\documentclass[aps,twocolumn,noshowpacs,nobalancelastpage,tightenlines,floatfix]{revtex4}
\usepackage{amssymb}
\usepackage{graphicx}
\usepackage[dvips]{epsfig}
\usepackage{dcolumn}
\usepackage{bm}
\usepackage{amsmath}

\begin{document}

\title{Functional Quantum Nodes for Entanglement Distribution\\
over Scalable Quantum Networks}
\author{Chin-Wen Chou, Julien Laurat, Hui Deng, Kyung Soo Choi, Hugues de
Riedmatten\footnote{Current address: Group of Applied Physics, University of Geneva, Geneva, Switzerland
}, Daniel Felinto\footnote{Current address: Departamento de F{\'i}sica, Universidade Federal de Pernambuco, Recife-PE, 50670-901, Brazil}, H. Jeff Kimble }
\affiliation{Norman Bridge Laboratory of Physics 12-33, California Institute of
Technology, Pasadena, California 91125, USA}
\date{\today}

\begin{abstract}
We 
 demonstrate  entanglement
distribution between two remote quantum nodes   located 3 meters
apart~\cite{SciExp}. This distribution involves the asynchronous preparation of
two  
pairs of atomic memories and the coherent mapping of stored atomic
states into light fields in an effective state of near maximum
polarization entanglement. Entanglement is verified by way of the
measured violation of a Bell inequality, and can be used for
communication protocols such as quantum cryptography.  
   The
demonstrated quantum nodes and channels can be used as segments of a
quantum repeater, providing an essential tool for robust
long-distance quantum communication.
\end{abstract}
\maketitle

In 
  quantum information  science~\cite{zoller05}, distribution of
entanglement over quantum networks is a critical requirement, including for
metrology~\cite{giovannetti04}, quantum computation~\cite{cirac97,duan04}
and communication~\cite{cirac97,briegel98}. Quantum networks are composed of
quantum nodes for processing and storing quantum states, and quantum
channels that link the nodes.   
Significant advances have
been made with diverse systems towards the realization of such networks,
including ions~\cite{blinov04}, single trapped atoms in free space~\cite{volz06,beugnon06} and in cavities~\cite{boozer07},
and atomic ensembles in the regime of continuous variables~\cite%
{polzik-tport}.

An approach of particular importance has been the seminal work of Duan,
Lukin, Cirac, and Zoller (DLCZ) for the realization of quantum
networks based upon entanglement between single photons and collective
excitations in atomic ensembles~\cite{duan01}. Critical experimental
capabilities have been achieved, beginning with the generation of
nonclassical fields~\cite{kuzmich03,vanderwal03} with controlled waveforms
\cite{balic05} and extending to the creation and retrieval of single
collective excitations~\cite%
{chou04,Eisaman05,Chaneliere05} with high efficiency~\cite%
{laurat06,thompson06}. Heralded entanglement with quantum memory, which is
the cornerstone of networks with efficient scaling, was achieved between two
ensembles  
~\cite{chou05}. More recently, conditional control of the
quantum states of a single ensemble \cite%
{deRiedmatten06,Matsukevich06b,chen06} and of two distant ensembles~\cite%
{felinto06} has also been implemented, as are likewise required for
the scalability of quantum networks based upon probabilistic
protocols.

\begin{figure*}[tbph]
\includegraphics[width=6in]{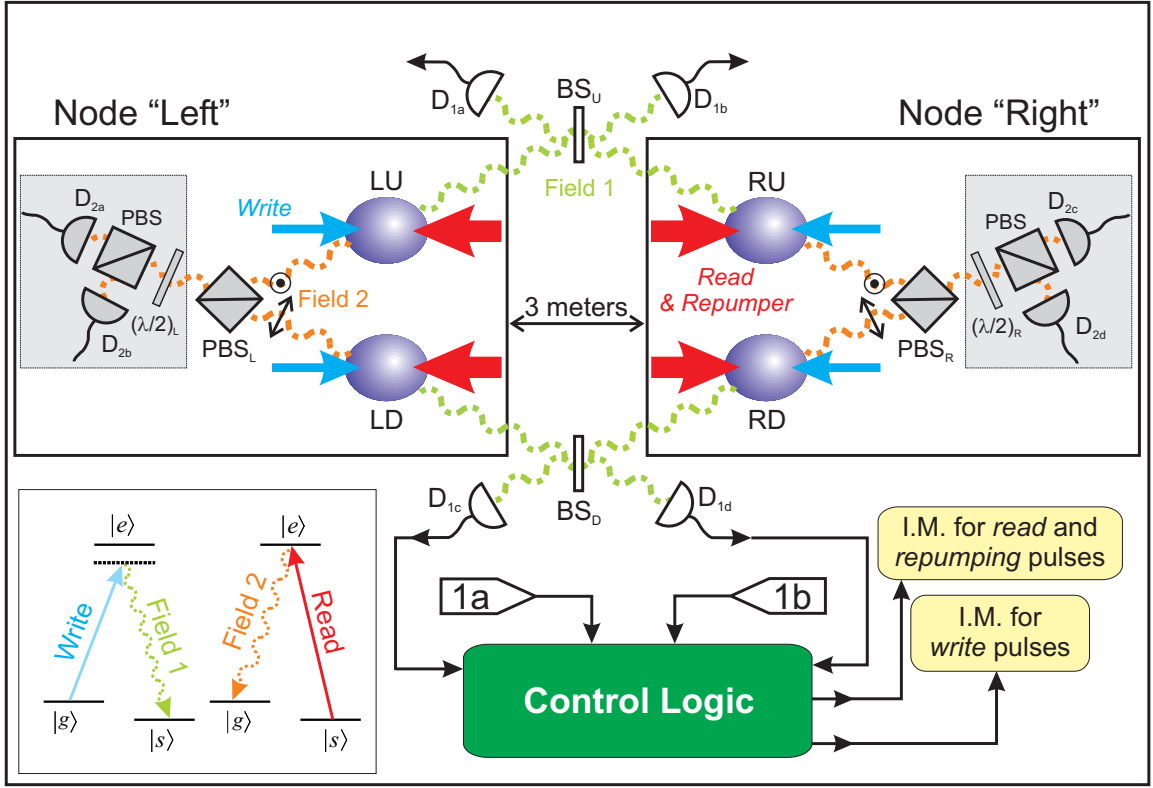}
\caption{Setup for distributing entanglement between two quantum
nodes $(L,R)$ separated by 3 meters. The inset shows the relevant
atomic levels for the $6S_{1/2}\rightarrow 6P_{3/2}$ transition in
atomic Cesium, as well as the associated light fields. The ensembles
are initially prepared in $|g\rangle $. Weak write pulses then
induce spontaneous Raman transitions $|g\rangle \rightarrow
|e\rangle \rightarrow |s\rangle $, resulting in the emission of
anti-Stokes fields (fields $1$) near the $|e\rangle \rightarrow
|s\rangle $ transition along with the storage of collective
excitations in the form of spin-flips shared among the
atoms~\protect\cite{duan01}. 
With this setup, a photo-detection event at either detector $D_{1a}$ or $D_{1b}$ ($D_{1c}$ or $D_{1d}$)
heralds
entanglement between the collective excitation in $LU$ and $RU$ ($LD$ and $RD$) \protect\cite%
{chou05} (see text). 
$BS_{U}$ and $BS_{D}$ are implemented using two orthogonal polarizations in one fiber beam splitter, yielding excellent relative path stability. A heralding detection event
triggers the control
logic to gate off the light pulses going to the corresponding ensemble pair (%
$U$ or $D$) by controlling the intensity modulators (\textit{IM}). The
atomic state is thus stored while waiting for the second ensemble pair to be
prepared. After both pairs of ensembles $U,D$ are entangled, the control
logic releases strong read pulses to map the states of the atoms to Stokes
fields $2$ via $|s\rangle \rightarrow |e\rangle \rightarrow |g\rangle $.
Fields $2_{LU}$ and $2_{LD}$ ($2_{RU}$ and $2_{RD}$) are combined with
orthogonal polarizations on the polarizing beam splitter \textit{PBS}$_{L}$ (%
\textit{PBS}$_{R}$) to yield field $2_{L}$ ($2_{R}$). If only coincidences
between fields $2_{L}$ and $2_{R}$ are registered, the state is effectively
equivalent to a polarization maximally entangled state. $(\protect\lambda %
/2)_{L,R}$: rotatable half-wave plates.}
\label{crypto_setup}
\end{figure*}

Our interest is to develop the physical resources that enable
quantum repeaters~\cite{briegel98},   
thereby  allowing  entanglement-based
quantum communication tasks over quantum networks on distance scales
much larger than set by the attenuation length of optical fibers,
including quantum cryptography~\cite{ekert91}. For this purpose,   
heralded number state entanglement  
\cite{chou05} between two
remote atomic ensembles is not directly applicable. Instead,  
DLCZ proposed to  use 
pairs of ensembles
$(U_{i},D_{i})$ at each quantum node $i$, with the sets of ensembles
$\{U_{i}\},\{D_{i}\}$ separately linked in parallel chains across
the network~\cite{duan01}. Relative to the state of the art in Ref.
\cite{chou05}, the DLCZ protocol requires the  
capability for the independent control of pairs of entangled
ensembles between two nodes.

In  our  
experiment,  
 we have created, addressed, and
controlled pairs of atomic ensembles at each of two quantum nodes,
thereby  
demonstrating  entanglement distribution in a form suitable both for quantum network  architectures and for entanglement-based
quantum communication schemes. 
Specifically, two
pairs of remote ensembles 
at two nodes are each prepared
in an entangled state~\cite{chou05}, in a heralded and asynchronous
fashion~\cite{felinto06}, thanks to the conditional control of the
quantum memories. After a signal heralding that the two chains are
prepared in the desired state, the states of the ensembles are
coherently transferred to propagating fields locally at the two
nodes. The fields are arranged such that they effectively contain
two photons, one at each node, whose polarizations are entangled.
The entanglement between the two nodes is verified by the violation
of a Bell inequality. The effective polarization entangled state,
created with favorable scaling behavior, 
is thereby compatible with
entanglement-based quantum communication protocols~\cite{duan01}.

The architecture for our experiment is shown in Figure~1.
Each quantum node, $L$ (left) and $R$ (right), consists of two atomic
ensembles, $U$ (up) and $D$ (down), or four ensembles altogether, namely ($LU
$, $LD$) and ($RU$, $RD$), respectively. Each pair is first prepared in an
entangled state, where one excitation is shared coherently, by using a pair
of coherent weak write pulses to induce spontaneous Raman
transitions $|g\rangle \rightarrow |e\rangle \rightarrow |s\rangle $, as
illustrated in the inset in Fig.~1.  The Raman fields ($%
1_{LU}$,$1_{RU}$) from ($LU$, $RU$) are combined at the 50-50
beamsplitter \textit{BS}$_{U}$, with the resulting fields directed
to single-photon detectors. A photoelectric detection event in
either detector heralds that the two ensembles are prepared. The
remote\ pair of $D$\ ensembles, ($LD $, $RD$), is prepared in an
analogous fashion.

Conditioned upon the preparation of both ensemble pairs ($LU$, $LD$)
and ($RU$, $RD$), a set of  read  pulses are triggered to map
the stored atomic excitations into propagating Stokes fields in
well-defined spatial modes via $|s\rangle \rightarrow |e\rangle
\rightarrow |g\rangle $ by way of a collective enhancement
\cite{duan01} (see inset in Fig.~1). This generates
a set of four fields denoted by ($2_{LU}$,$2_{RU}$) for ensembles
($LU$, $RU$) and ($2_{LD} $,$2_{RD}$) for ensembles ($LD$, $RD$). In
the ideal case and neglecting higher-order terms, this mapping results in a quantum state for the
fields $2$ given by
\begin{widetext}
\begin{eqnarray}
|\psi _{2_{LU},2_{RU},2_{LD},2_{RD}}\rangle
=\frac{1}{2}\Big(|0\rangle _{2_{LU}}|1\rangle _{2_{RU}}\pm e^{i\eta
_{U}}|1\rangle _{2_{LU}}|0\rangle _{2_{RU}}\Big)_{U}\otimes
\Big(|0\rangle _{2_{LD}}|1\rangle _{2_{RD}}\pm e^{i\eta
_{D}}|1\rangle _{2_{LD}}|0\rangle _{2_{RD}}\Big)_{D}\text{ .}
\label{psi-all}
\end{eqnarray}
\end{widetext}
Here, $|n\rangle _{x}$ is the $n$-photon state for mode $x$, where
$x\in \{2_{LU},2_{RU},2_{LD},2_{RD}\}$, and $\eta _{U}$ ($\eta
_{D}$) is the relative phase resulting from the writing and reading
processes for the $U$ ($D$) pair of ensembles~\cite{chou05}.\ The
$\pm $ signs for the conditional states $U,D$ result from the
unitarity of the transformation by the beamsplitters ($BS_{U}$,
$BS_{D}$). The extension of Eq.~\eqref{psi-all} to incorporate various
nonidealities is given in the Appendix.

Apart from an overall phase, the state $|\psi
_{2_{LU},2_{RU},2_{LD},2_{RD}}\rangle $\ can be rewritten as follows:
\begin{widetext}
\begin{eqnarray}
|\psi _{2_{LU},2_{RU},2_{LD},2_{RD}}\rangle
&=&\frac{1}{2}\Big[e^{-i\eta _{D}}|1\rangle _{2_{RU}}|1\rangle
_{2_{RD}}|vac\rangle _{2_{L}}\pm e^{i\eta _{U}}|1\rangle
_{2_{LU}}|1\rangle _{2_{LD}}|vac\rangle
_{2_{R}}\nonumber\\&&\pm \big(|0\rangle
_{2_{LU}}|1\rangle _{2_{LD}}|0\rangle _{2_{RD}}|1\rangle
_{2_{RU}}\pm e^{i(\eta _{U}-\eta _{D})}|1\rangle
_{2_{LU}}|0\rangle _{2_{LD}}|1\rangle _{2_{RD}}|0\rangle
_{2_{RU}}\big)\Big], \nonumber\\\label{psi-all-2}
\end{eqnarray}
\end{widetext}
where $|vac\rangle _{2_{i}}$ denotes $|0\rangle _{2_{iU}}|0\rangle
_{2_{iD}}$. If only coincidences between both nodes $L,R$ are
registered, the first two terms (i.e., with $e^{-i\eta
_{D}},e^{i\eta _{U}}$) do not contribute. Hence, as noted by DLCZ,
excluding such cases leads to an effective density matrix equivalent
to the one for a
maximally entangled state of the form of the last term in Eq.~\eqref{psi-all-2}%
. Significantly, the absolute phases $\eta _{U}$ and $\eta _{D}$ do not need
to be independently stabilized. Only the relative phase $\eta =\eta
_{U}-\eta _{D}$ must be kept constant, leading to $1/2$ unit of entanglement for two quantum bits (i.e., 1/2 ebit).

The experimental demonstration of this architecture for implementing
the DLCZ protocol relies critically on the ability to carry out
efficient parallel preparation of the ($LU$, $RU$) and ($LD$, $RD$)
ensemble pairs, as well as to stabilize the relative phase $\eta $.
The first requirement
is achieved by the use of real-time control, as described in Ref. \cite%
{felinto06} in a simpler case. As illustrated in Fig.
1, here we implement control logic that monitors
the outputs of field 1 detectors. A detection event at either pair
triggers electro-optic intensity modulators (\textit{IM})\ that gate
off all laser pulses going to the corresponding pair of ensembles,
thereby storing the associated state. Upon receipt of signals
heralding that the two pairs of ensembles ($LU$, $RU$) and ($LD$,
$RD$) have both been independently prepared, the control logic
triggers the retrieval of the stored states by simultaneously
sending a strong  read  pulse into each of the four
ensembles. Relative to the case where no logic is implemented, a
19-fold enhancement is obtained in the probability to generate this
overall state from the four ensembles (see Appendix).

The second requirement, for stability of the relative phase $\eta $,
could be accomplished by active stabilization of each individual
phase $\eta _{U},\eta _{D}$, as in Ref.~\cite{chou05}. Instead of
implementing this challenging technical task (which ultimately would
have to be extended across longer chains of ensembles), our setup
exploits the passive stability between two independent polarizations
propagating in a single interferometer to prepare the two ensemble
pairs   
\cite{chouThesis}. No active phase
stabilization is thus required. In practice, we find that the
passive stability of our system is sufficient for operation
overnight without adjustment. Additionally, we implement a procedure
that deterministically sets the relative phase $\eta $ to zero.

We also extend the original DLCZ protocol by, as illustrated in Fig.~1, 
combining fields ($2_{LU}$, $2_{LD}$) and ($2_{RU}$, $%
2_{RD}$) with orthogonal polarizations on polarizing beam splitters \textit{%
PBS}$_{L}$ and \textit{PBS}$_{R}$ to yield fields $2_{L}$ and
$2_{R}$, respectively. The polarization encoding opens the
possibility of performing additional entanglement purification and
thus superior scalability~\cite{jiang06,chen06a}. In the ideal case, the
resulting state is now effectively equivalent to a maximally
entangled state for the polarization of two photons

\begin{equation}
|\psi _{2_{L},2_{R}}^{\pm }\rangle _{eff}\propto |H_{2_{L}}\rangle
|V_{2_{R}}\rangle \pm e^{i\eta }|V_{2_{L}}\rangle |H_{2_{R}}\rangle \text{ ,}
\label{P_ent}
\end{equation}%
where $|H\rangle $ ($|V\rangle $) stands for the state of a single
photon with
horizontal (vertical) polarization. The sign of the superposition in Eq.~\eqref%
{P_ent} is inherited from Eq.~\eqref{psi-all} and is determined by the
particular pair of heralding signals recorded by($D_{1a}$,$D_{1b}$)
and ($D_{1c}$,$D_{1d} $). The entanglement in the polarization basis
is well-suited for entanglement-based quantum cryptography
\cite{duan01,ekert91}, including security verification by way of the
violation of a Bell inequality, as well as for quantum teleportation
\cite{duan01}.

As a first step to investigate the joint states of the atomic ensembles, we
record photoelectric counting events for the ensemble pairs $(LU,RU)$ and $%
(LD,RD)$ by setting the angles for the half-wave plates $(\frac{\lambda }{2}%
)_{L,R}$\ shown in Fig.~1 to $0^{\circ }$ such that photons
reaching detectors $D_{2b}$ and $D_{2d}$ ($D_{2a}$ and $D_{2c}$) come only
from the ensemble pair $U$ ($D$). Conditioned upon detection events at $%
D_{1a}$ or $D_{1b}$ ($D_{1c}$ or $D_{1d}$), we estimate the probability that
each ensemble pair $U,D$ contains only a single, shared excitation as
compared to the probability for two excitations by way of the associated
photoelectric statistics. In quantitative terms, we determine the ratio \cite%
{chou05}

\begin{equation}
h_{X}^{(2)}\equiv \frac{p_{X,11}}{p_{X,10}p_{X,01}}\text{ ,}
\end{equation}%
where $p_{X,mn}$ are the probabilities to register $m$ photo-detection
events in mode $2_{LX}$ and $n$ events in mode $2_{RX}$ ($X=\{U,D\}$)
conditioned on a detection event at $D_{1}$. A necessary condition for the
two ensembles ($LX$, $RX$) to be entangled is that $h_{X}^{(2)}<1$, where $%
h_{X}^{(2)}=1$ corresponds to the case of independent (unentangled)
coherent states for the two fields~\cite{chou05}. Fig.~2
shows the measured $h_{X}^{(2)}$ versus the duration $\tau _{M}$
that the state is stored before retrieval. For both $U$ and $D$
pairs, $h^{(2)}$ remains well below unity for storage times $\tau
_{M}\lesssim 10$~$\mu s$. For the $U$ pair, the solid line in Fig.~2 provides a fit by the simple expression $h^{(2)}=1-A\exp
{\left( -(\tau _{M}/\tau )^{2}\right) }$. The fit gives $A=0.94\pm
0.01$ and $\tau =22\pm 2$ $\mu s$, providing an estimate of a coherence
time for our system. A principal cause for decoherence is an
inhomogeneous broadening of the ground state levels by residual
magnetic fields~\cite{felinto05}. The characterization of the time dependence of $h^{(2)}$ constitutes an important benchmark of our system, as will be clarified shortly (also see Appendix).

\begin{figure}[t]
\vspace{-8mm}\centering
\includegraphics[width=0.99\columnwidth]{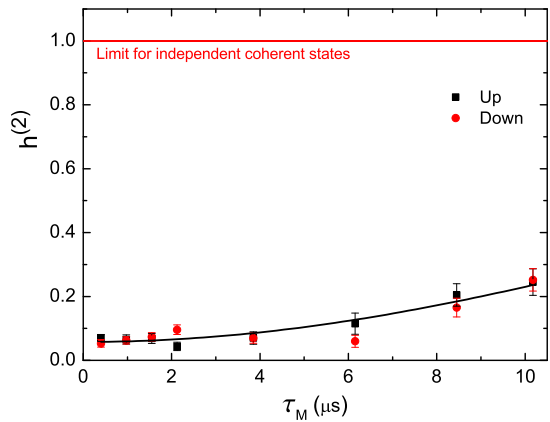} \noindent
 \vspace{-9mm}\caption{Suppression $h^{(2)}$ of the probabilities for
each ensemble to emit two photons compared to the product of the
probabilities that only one photon is emitted, as a function of the
duration $\protect\tau _{M}$ that the state is stored before
retrieval. The solid line gives a fit for the $U$ pair. Error bars
indicate statistical errors.} \label{h2}
\end{figure}

We next measure the correlation function $E(\theta _{L},\theta
_{R})$ defined by
\begin{equation}
E(\theta _{L},\theta _{R})=\frac{C_{ac}+C_{bd}-C_{ad}-C_{bc}}{%
C_{ac}+C_{bd}+C_{ad}+C_{bc}}\text{ .}
\end{equation}
Here, $C_{jk}$ gives the rates of coincidences between detectors
$D_{2j}$ and $D_{2k}$ for fields $2$, where $j,k\in \{a,b,c,d\}$,
conditioned upon heralding events at detectors $D_{1a},D_{1b}$ and
$D_{1c},D_{1d}$ from
fields $1$. The angles of the two half-wave plates $(\frac{\lambda }{2}%
)_{L,R}$ are set at $\theta _{L}/2$ and $\theta _{R}/2$,
respectively. As stated before, the capability to store the state
heralded in one pair of ensembles and then to wait for the other
pair to be prepared significantly improves the various coincidence
rates $C_{jk}$ by a factor that increases with the duration $\tau
_{M}$ that a state can be preserved~\cite{felinto06} (also see Appendix).

\begin{figure}[b]
\vspace{-7mm}\centering
\includegraphics[width=0.98\columnwidth]{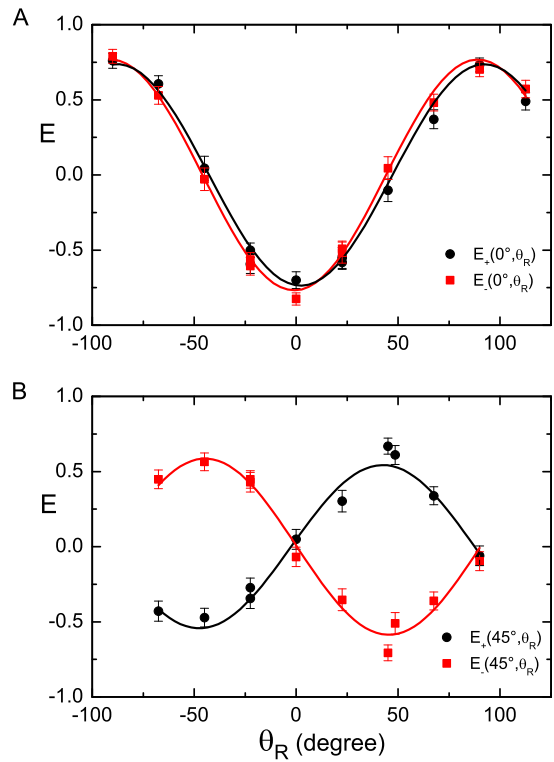}
\noindent \hrulefill
\vspace{-8mm}\caption{Measured correlation function $E(\protect\theta _{L},\protect\theta %
_{R})$ as a function of $\protect\theta _{R}$ with $\protect\theta
_{L}$ fixed at  A:  $0^{\circ }$ and  B:  $45^{\circ }$. The
excitation probabilities for the ensembles are increased by $\sim
1.5$ times relative to Fig. 2, with each point taken for 30 minutes
at typically 400/hour coincidence rate for each fringe. Error bars
indicate statistical errors.} \label{fringes}
\end{figure}

Fig.~3 displays the correlation function $E$ as a function of $%
\theta _{R}$, with $\theta _{L}=0^{\circ }$ in  A  and $45^{\circ }$
in  B . Relative to Fig.~2, these data are taken with
increased excitation probability (higher write power) to validate
the phase stability of the system, which is evidently good.
Moreover, these four-fold coincidence fringes in
Fig.~3A  provide a further verification that
predominantly one excitation is shared between a pair of ensembles.
The analysis provided in the Appendix
with the measured cumulative $h^{(2)}$ parameter for this set of data, $%
h^{(2)}=0.12\pm$   $0.02$  , predicts a visibility of $V=78\pm 3\%$ in good
agreement with the experimentally determined $V\cong 75\%$. Finally, the
fact that one of the fringes is inverted with respect to the other in Fig.~%
3B corresponds to the two possible signs in
Eq.~\eqref{P_ent}. As for $\theta _{L}=45^{\circ }$ the
measurement is sensitive to the square of the overlap $\xi $ of
photon wavepackets for fields $2_{U,D}$, we may infer $\xi
_{U,D}\simeq 0.85$ from the reduced fringe visibility ($V\cong
55\%$) in Fig.~3B relative to  A , if all the
reduction is attributed to a nonideal overlap. An independent
experiment for two-photon
interference in this setup has shown an overlap $\xi \simeq 0.90$ (see Appendix), which confirms that the reduction can be principally
attributed to the non-ideal overlap. Other possible causes include
imperfect phase alignment $\eta \neq 0$ and imbalance of the
effective state coefficients (see Appendix).

\begin{figure*}[tbph]
\vspace{-8mm}\centering
\includegraphics[width=5in]{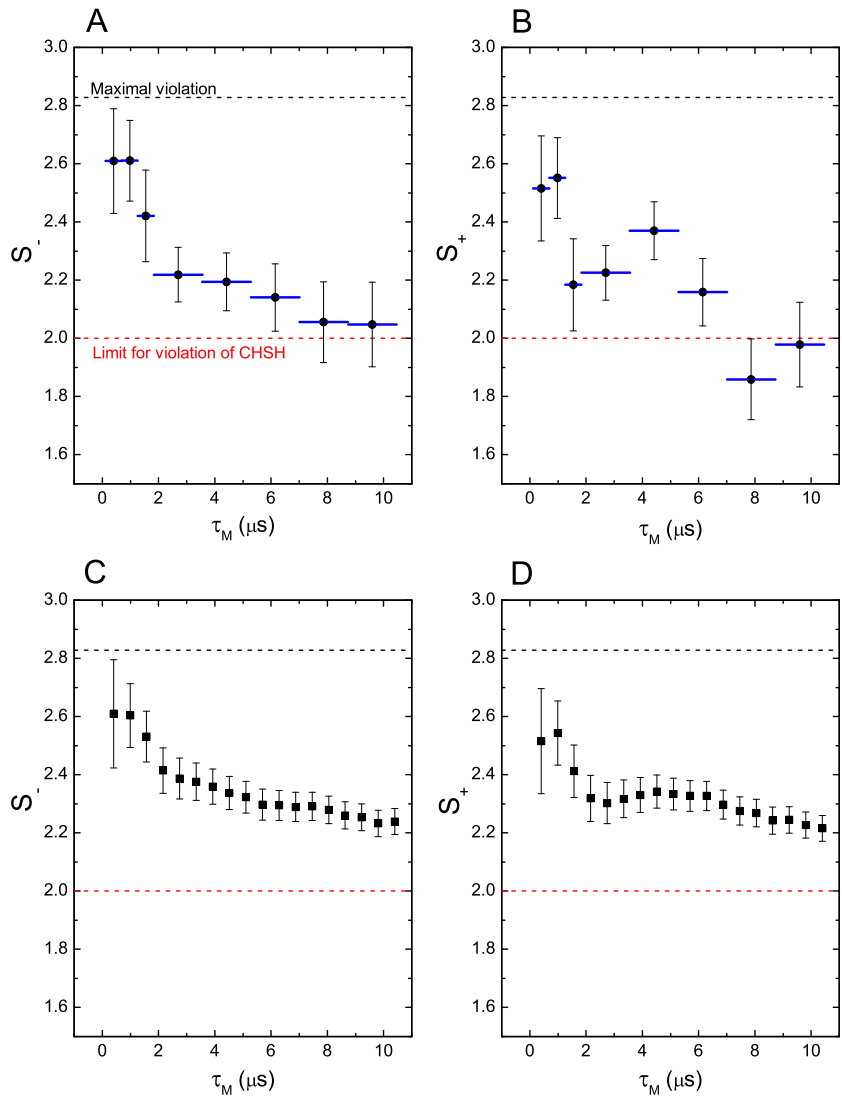}
\vspace{-8mm}\caption{Measured CHSH parameters $S_{\pm }$, for the
two possible effective states in Eq.~\eqref{P_ent}, as functions of
duration $\protect\tau _{M}$ for which the first ensemble pair holds
the prepared state. The excitation
probabilities are kept low for high correlation (as in Fig.~2).  A  and  B: binned data. The horizontal thick lines indicate the size
of the bins used. C and D: cumulative data. The coincidence rate for
these measurements is about 150/hour for each effective state. Error bars
indicate statistical errors.}
\label{S}
\end{figure*}

With the measurements from Figs.~2,~3 in hand, we
verify
entanglement unambiguously by way of the violation of a Bell inequality \cite%
{clauser78}. For this purpose, we choose the canonical values,
$\theta _{L}=\{0^{\circ },45^{\circ }\}$ and $\theta
_{R}=\{22.5^{\circ },-22.5^{\circ }\}$, and construct the CHSH
parameters
\begin{eqnarray}
S_{+} &\equiv &|E(0^{\circ },22.5^{\circ })+E(0^{\circ
},-22.5^{\circ
})\qquad \qquad  \notag \\
&&+E(45^{\circ },-22.5^{\circ })-E(45^{\circ },22.5^{\circ })|  \label{S+-} \\
S_{-} &\equiv &|E(0^{\circ },22.5^{\circ })+E(0^{\circ
},-22.5^{\circ })
\notag \\
&&+E(45^{\circ },22.5^{\circ })-E(45^{\circ },-22.5^{\circ })|
\end{eqnarray}%
for the two effective states $|\psi _{2_{L},2_{R}}^{\pm }\rangle
_{eff}$ in Eq.~\eqref{P_ent}. For local, realistic hidden variable
theories, $S_{\pm }\leq 2$~\cite{clauser78}. Fig.~4 shows the
CHSH parameters $S_{\pm }$ as functions of the duration $\tau _{M}$
up to which one pair of ensembles holds the prepared state, in the
excitation regime of Fig.~2. As shown in the Appendix, the requirements for minimization of
higher order terms are much more stringent in this
experiment with four ensembles than with simpler configurations \cite%
{deRiedmatten06}.

Panels  A  and  B  in Fig.~4  give the results for our
measurements of $S_{\pm }$ with binned data. Each point corresponds
to the violation obtained for states generated at $\tau _{M}\pm
\Delta \tau _{M}/2$ ($\Delta \tau _{M}$ marked by the thick
horizontal lines in Fig.~4). Strong violations are obtained
for short memory times, with for instance $S_{+}=2.55\pm 0.14>2$ and
$S_{-}=2.61\pm 0.13>2 $ for the second bin, demonstrating the
presence of entanglement between fields $2_{L}$ and $2_{R}$.
Therefore, these fields can be exploited to perform
entanglement-based quantum communication protocols, such as quantum
key distribution with, at minimum, security against individual attacks \cite%
{duan01,fuchs97}.

As it can be seen in Fig.~4, the violation decreases with increasing $%
\tau _{M}$. 
The decay is largely due to the time-varying behavior of $h^{(2)}$ (Fig.~2 and Appendix). In addition to this
decay, the $S_{+}$ parameter exhibits modulation with $\tau _{M}$. We have explored different
models for the time dependence of the CHSH parameters, but thus far
have found no satisfactory agreement between model calculations and
measurements. Nevertheless, the density matrix for the ensemble over
the full memory time is potentially useful for tasks such as
entanglement connection, as shown by panels C and D where
cumulative data are given. Each point at memory time $\tau _{M}$
gives the violation obtained by taking into account all the states
generated from $0$ to $\tau _{M}$. Overall significant violations
are obtained, namely $S_{+}=2.21\pm 0.04>2$ and $S_{-}=2.24\pm
0.04>2$ at $\tau _{M}\sim 10\mu s$.

In our experiment, we are able to generate excitation-number entangled states between remote
locations, which are well suited for scaling purposes, and, with real-time control, to operate them as if they were effectively polarization entangled states, which are appropriate for quantum communication
applications, such as quantum cryptography. Measurements of the suppression $h^{(2)}$ of two-excitation components versus storage time demonstrates explicitly the major source that causes the extracted polarization entanglement to decay, emphasizing the critical role of multi-excitation events in the experiments aiming for a scalable quantum network. The present scheme, which constitutes a functional segment of a quantum repeater in terms of quantum state encoding and channel control, allows the distribution of entanglement between two quantum nodes. But the extension of our work to longer chains involving many segments becomes more complicated, and out of reach for any current system. For long-distance communication, the first quantity to improve is the coherence time of the memory. Better cancellation of the residual magnetic fields and switching to new trap schemes should improve this parameter to $\sim$ 0.1 sec by employing an optical trap~\cite{felinto05}, thereby increasing the rate of preparing the ensembles in the state of Eq.~\eqref{psi-all} to $\sim$ 100~Hz.  The second 
challenge which would immediately appear in an extended chain would be the increase of the multi-excitation probability with the connection stages. Recently, Ref.~\cite{jiang06} has theoretically demonstrated the prevention of such growth in a similar setup, but its full scalability still requires very high retrieval and detection efficiency, and photon-number resolving detectors. These two points clearly show that the quest of scalable quantum networks is still a theoretical and experimental challenge. The availability of our first functional segment opens the way for fruitful investigations. \bigskip

\appendix

\section{Experimental details}

Ensembles ($LU$, $LD$) are pencil-shaped groups of cold Cesium atoms
in a magneto-optical trap (MOT) while ensembles ($RU$, $RD$) are in
another MOT, 3 meters away. $\{|g\rangle ,|s\rangle ,|e\rangle \}$
correspond to the hyperfine levels $\{|6S_{1/2},F=4\rangle
,|6S_{1/2},F=3\rangle ,|6P_{3/2},F^{\prime }=4\rangle \}$,
respectively. In each MOT, the ensembles $U,D$ are separated by 1 mm
by way of birefringent beam displacers \cite{chouThesis}. The MOT is
formed at a repetition rate of 40 Hz. In each cycle, the MOT is
loaded for 18 ms, after which the magnetic field is quickly switched
off. The trapping beams are turned off 3 ms after the magnetic
field, while the repumping beam stays on for another
100 $\mu $s before being switched off in order to prepare the atoms in the $%
F=4$ ground state $|g\rangle$. 3.4 ms after the magnetic field is
turned off, trials of the protocol (each consisting of successive
write, read, and repumping pulses) are repeated with 575 ns period
for 3.4 ms. In each trial, the write
pulse is $\approx $ 30 ns in duration and 10 MHz red-detuned from the $%
|g\rangle \rightarrow |e\rangle $ transition. The read pulse and the
repumping pulse are both derived from the read beam (resonant with the $%
|s\rangle \rightarrow |e\rangle $ transition) with 30 ns and 75 ns
duration, respectively. The read pulse is closely followed by the
repumping pulse. The read pulse is delayed $\approx $ 400 ns after
the write pulse, leaving time for the control logic to gate it off,
along with the subsequent pulses. Independent phase stability
measurements show that the phase $\eta$ drifts in a negligible way,
($\pi/30$) over $500\mu s$ corresponding to 870 trials. Some other
parameters of the experiments are calibrated and listed in
table~\ref{table}.

\begin{table*}[htpb]
\centering \caption{Noise and Efficiencies} \label{table}
\vspace{0.1in}
\begin{tabular}{|c|c|c|}
\hline
\qquad\qquad\qquad\qquad\qquad\qquad\qquad\qquad\qquad\qquad\qquad&
\quad\qquad\qquad U \qquad\qquad\quad\quad & \qquad\quad\qquad D
\qquad\qquad\quad\quad\\ \hline Field 1 dark count rate & $\sim$ 10
Hz & $\sim$ 10 Hz \\ \hline Field 2 dark count rate & $\sim$ 100 Hz
& $\sim$ 100 Hz \\ \hline Overall retrieval efficiency $p_c$ &
$6.4\%\pm 0.5\% $ & $8.0\%\pm 0.5\% $ \\ \hline Field 2 propagation
loss & $68\pm 5\%$ & $68\pm 5\%$ \\ \hline Field 2 photon detection
efficiency & $50\pm 5\%$ & $50\pm 5\%$ \\ \hline
\end{tabular}
\end{table*}

\section{Fringe visibility as a function of $h^{(2)}$}

Let us consider that the two pairs of ensembles, \textit{U} and \textit{D},
have been prepared by heralded detections at $D_{1a},D_{1b}$ and $%
D_{1c},D_{1d}$. Denote by $p_{10}$, $p_{01}$, and $p_{11}$ the probability $%
p_{ij}$ to register $i$ photodetection events in field $2_{LU}$ and $j$ in
field $2_{RU}$ after firing the read pulses. We will assume, for simplicity,
the various $p_{ij}$ are the same for both pairs of ensembles. For each of
them, the suppression of the two-photon events relative to the square of the
probability for single-photon events is characterized by the parameter $%
h^{(2)}$\cite{chou05}:
\begin{equation}
h^{(2)}=\frac{p_{11}}{p_{10}p_{01}}\,.
\end{equation}

We next relate $h^{(2)}$\ to the maximal $C_{max}$ and minimal $C_{min}$
coincidence probabilities between various output ports of the detection
polarizing beamsplitters ($PBS$) for the left and right nodes at detectors $%
D_{2a},D_{2b}$ and $D_{2c},D_{2d}$ (see Fig. 1 of the main text).
Consider, for example, the transmitted ports of the $PBS$ at the $L,R$
detectors for the case that the left node has the half-wave plate $(\frac{%
\lambda }{2})_{L}$ set to $0^{\circ }$. In this case, fields $2_{LU}$ and $%
2_{LD}$ are detected independently, with field $2_{LD}$ transmitted at the $%
PBS$. On one hand, $C_{max}$ is obtained for crossed polarizers (i.e., $(%
\frac{\lambda }{2})_{R}$ set to $45^{\circ }$ at the right node, with then
field $2_{RU}$ transmitted) and is given to lowest order by:
\begin{equation}
C_{max}=p_{10}p_{01}.  \label{max}
\end{equation}%
This term corresponds to the case where only a single excitation is
distributed in each pair, and each retrieved photon is detected from a
transmitted port on each side $L,R$.

On the other hand, the minimum coincidence probability $C_{min}$ is obtained
for parallel polarizers (i.e., $(\frac{\lambda }{2})_{R}=0^{\circ }$ at the
right node, with then field $2_{RD}$ transmitted) and can be written as:
\begin{equation}
C_{min}=p_{11}\text{.}  \label{min}
\end{equation}%
This term corresponds to coincidences due to photons coming from the same
pair of ensembles. The smaller is the excitation probability, the smaller is
this background term.

Taking Eqs.~\eqref{max} and~\eqref{min} into account, we find that the
visibility for the number of coincidences as a function of the right
polarizer angle (i.e., the angle for $(\frac{\lambda }{2})_{R}$) is given by:%
\begin{equation}
V=\frac{C_{max}-C_{min}}{C_{max}+C_{min}}=\frac{1-h^{(2)}}{1+h^{(2)}}\text{ .%
}
\end{equation}%
Assuming that the visibility is the same in each basis, we then find a CHSH
parameter $S$ equal to \cite{Marcikic04}:
\begin{equation}
S=2\sqrt{2}\,V=2\sqrt{2}\,\frac{1-h^{(2)}}{1+h^{(2)}}\text{ .}
\end{equation}

A minimal value $h_{min}^{(2)}=0.17$ is thus required to violate the CHSH
inequality $S<2$ in the absence of any imperfections except the intrinsic
two-photon component. This value underlines that this experiment is much
more stringent in terms of minimization of high-order terms than previously
reported setups. For example, in Ref. \cite{deRiedmatten06}, where
entanglement between a photon and a stored excitation is reported, a value
of $h^{(2)}$ equal to 0.68 was sufficient to violate the inequality. The
dramatic improvement reported recently by different groups for the quality
of the photon pairs emitted by an atomic ensemble was thus an enabling step
for the practical realization of such a more elaborate procedure involving a
total of 4 ensembles reported in the main text.

\section{Two-photon interference and inferred overlaps}

For a non-perfect overlap $\xi $ of the field-2 photon wavepackets, the
visibility of the fringes in the $45^{\circ }$ basis is decreased by a
factor $\xi ^{2}$. This overlap can be determined by two-photon
interference, which is implemented by mixing the fields $2_{U}$ and $2_{D}$
on each side (Right and Left) by rotating the half-wave plates $(\frac{%
\lambda }{2})_{L},(\frac{\lambda }{2})_{R}$ by $22.5^{\circ }$. If the
single photon wavepackets are indistinguishable, no coincidences should be
observed. However, the two-photon component can lead to coincidences, which
reduce the visibility. Let us determine the expected visibility as a
function of the two-photon component by way of a simple model.

Consider $P_{n}$ the probability of finding $n$ photons in field 2, and
assume the various $P_{n}$ are the same for both ensembles involved. In the
ideal case where all ensembles have the same properties, the two-photon
suppression for each field 2 can also be characterized by the same $h^{(2)}$
parameter used before, which can be written here as:
\begin{equation}
h^{(2)}=\frac{2P_{2}}{P_{1}^{2}}.
\end{equation}

When the half-wave plates $(\frac{\lambda }{2})_{L},(\frac{\lambda }{2})_{R}$
are at $0^{\circ }$, the fields 2 are detected independently and the
probability $p_{max}$ to register coincidences is given by:
\begin{equation}
p_{max}=P_{1}^{2}.
\end{equation}

When the half-wave plates $(\frac{\lambda }{2})_{L},(\frac{\lambda
}{2})_{R}$ are rotated to $22.5^{\circ }$, if the two fields overlap
perfectly, the term with one photon in each input does not lead to
coincidences. If we denote by $\xi $ the overlap, the probability
$p_{min}$ to have one photon in each output is then:
\begin{equation}
p_{min}=\frac{(1-\xi^2
)}{2}P_{1}^{2}+\frac{P_{2}}{2}+\frac{P_{2}}{2}=[1-\xi^2
+h^{(2)}]\frac{P_{1}^{2}}{2}\text{ .}
\end{equation}

From these two probabilities, we find that the visibility of the dip in
coincidences can be written as:
\begin{equation}
V_{dip}=\frac{p_{max}-p_{min}}{p_{max}}=\frac{1+\xi^2-h^{(2)}}{2}\text{.}
\end{equation}

In our case, the measured visibility $V_{dip}$ is $85\pm 2\%$ for the left
node and $89\pm 2\%$ for the right one. The measured average $h^{(2)}$
parameter for this set of data is $0.09\pm 0.01$, which should lead in the
case of perfect overlap to visibilities $V_{\text{model}}=95.5\pm 0.5\%$.
From the measured visibilities and this simple model, we can then estimate
the overlaps: $\xi =0.89\pm 0.03$ for the left node and $\xi =0.93\pm 0.03$
for the right node.

\section{Decoherence time of the stored excitation}
\begin{figure}[htpb]
\centering
\includegraphics[width=0.7\columnwidth]{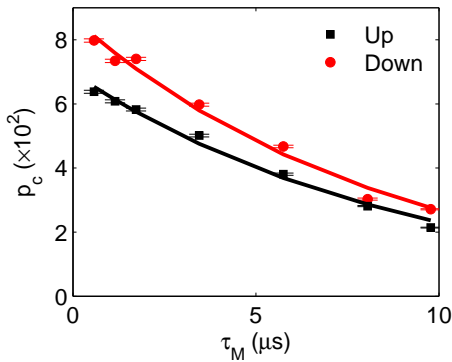}
\caption{Conditional probability $p_c$ of detecting one photon in a
field 2 for the U (black squares) and D (red circles) pairs, as a
function of the storage time $\tau_M$ of an excitation. The error
bars indicate statistical errors. The solid lines are fits using
Eq.~\eqref{eq:pc}.} \label{fig:pc}
\end{figure}

Residual magnetic fields, which lead to inhomogeneous broadening of
the ground states levels, is the major limiting factor of the
coherence time $\tau_c$ of the stored excitation
\cite{felinto05,deRiedmatten06}. Consequently, if we neglect dark
counts, the conditional retrieval efficiency $p_c=p_{01}+p_{10}$ is
expected to decay exponentially with the storage time $\tau_M$:
\begin{equation}
p_c = p^{0}_{c} \exp(-\frac{\tau_M}{\tau_c}). \label{eq:pc}
\end{equation}
Figure \ref{fig:pc} shows an independent measurement of $p_c$ vs.
$\tau_M$, with the $U$ and $D$ pairs separated. Fitting the data
with Eq.~\eqref{eq:pc} gives, for the U and D pairs respectively,
$p_{c}^{0}=7.0\%\pm 0.1\%$ and $8.7\%\pm 0.2\% $, and $\tau_c =
9.1\pm 0.6~\mu$s and $8.5\pm 0.5~\mu$s.

\begin{figure}[htpb]
\centering
\includegraphics[width=0.98\columnwidth]{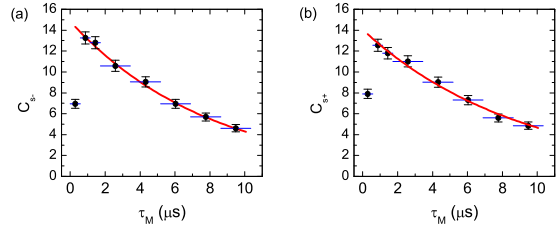}
\caption{The $\tau_M$ dependence of the total conditional count
rates $C_{S\pm}$ in the measurement of (a) $S_{+}$ and (b) $S_{-}$.
The horizontal thick lines indicate the size of the memory bin. The
error bars indicate statistical errors. The solid lines are fits
using Eq.~\eqref{eq:C}.} \label{fig:C}
\end{figure}

The decay of $p_c$ leads to a similar exponential decay of $C_{ij}$.
$C_{ij}$ ($i,j={a,b,c,d}$) are the coincidence count rates of two
field 2 photons conditioned on two heralding field 1 photons defined
before. Summing over all $C_{ij}$ used in calculating $S_{\pm}$, we
obtain the total coincidence count rates $C_{S\pm}$ for the
measurement of the Bell parameters $S_{+}$ and $S_{-}$.
$C_{S\pm}(\tau_M)$ corresponds to the probability distribution of
the $S_{\pm}(\tau_M)$, and is reflected in the statistical error
bars $\Delta S_{\pm}(\tau_M)$. The decay of $C_{S\pm}$ with $\tau_M$
is shown in Fig.~\ref{fig:C}. Fitting the data with exponential
functions:
\begin{equation}
C_{S\pm} = C^0_{S\pm}\exp(-\tau_M/\tau_{\pm}),\quad \tau_M>0,
\label{eq:C}
\end{equation}
gives $\tau_{+} =9.1\pm 0.4~\mu$s and $\tau_{-}= 8.1\pm 0.3~\mu$s,
in good agreement with $\tau_c$. Note that
$C^0_{S\pm}=2C_{S\pm}(\tau_M=0)$, since $C_{S\pm}(\tau=0)$ is
conditioned on two excitation in a same trial, while
$C_{S\pm}(\tau>0)$ is conditioned on two excitations created in two
different trials: the factor of two accounts for the two possible
orders of excitations ('U' then 'D' or 'D' then 'U').

\section{Conditional control and increase in generation rate}

\begin{figure}[htpb]
\centering
\includegraphics[width=0.95\columnwidth]{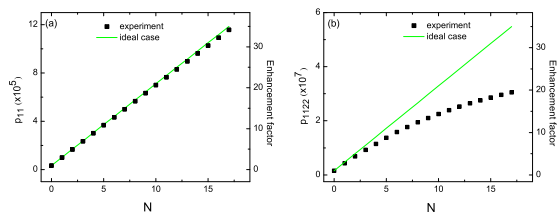}
\caption{Probabilities of coincidence detection as functions of the
number of trials $N$ for which the first prepared pair holds the
state. (a): measured probability $p_{11}$ of preparing the two
pairs. (b): measured probability $p_{1122}$ of detecting field 2
coincidences. The green solid line corresponds to the ideal case of
very long coherence time. Both panels give in addition to these
probabilities the enhancement factor obtained relative to the case
without conditional control.} \label{circuit}
\end{figure}

As demonstrated in Ref. \cite{felinto06}, the conditional control of remote
memories enables a large enhancement of coincidence rates relative to the
case where no logic is implemented. If the state prepared in one pair of
ensembles is held up to $N$ trials, the rate for preparing both pairs is
increased by a factor $(2N+1)$ for very low excitation probability \cite%
{felinto06}. Figure \ref{circuit}(a) gives the probability $p_{11}$ of
simultaneously preparing the two pairs. After 17 trials, an increase by a
factor 34 is obtained experimentally, close to the expected value of 35. The
gain in the probability $p_{1122}$ of generating the effective entangled
state is expected to be the same if the coherence time is long enough.
However, our finite coherence time results in a smaller increase of the
probability to detect field 2 coincidences. This increase is given in Fig. %
\ref{circuit}(b), with a comparison to the ideal case of very long
coherence time. A 19-fold enhancement is finally obtained. Let us
note that the different experimental rates can be obtained from
these probabilities times the number of trials per second ($\sim
2.36\times 10^{5}/$s).

\section{Correlation functions $E(0^\circ,\protect\theta_R)$, $E(45^\circ,%
\protect\theta_R)$ for the ideal effective state}

In practice, various imperfections lead to deviations from the ideal
effective state, Eq. (2) in the main text. We have developed a detailed
model relevant to our experiment, but consider here only a generic form.
Collective excitations are not shared with equal amplitudes between a pair
of ensembles because of nonidealities in the writing and heralding
processes. Likewise, the mapping of atomic states to states of field $2$ by
the read pulses is not ideal. Overall, these various imperfections lead to a
state $|\psi _{2_{LU},2_{RU},2_{LD},2_{RD}}\rangle $ for field $2$ given by
(neglecting multi-photon processes):
\begin{widetext}
\begin{eqnarray}
|\psi _{2_{LU},2_{RU},2_{LD},2_{RD}}\rangle  &=&\Big(\epsilon
_{RU}|0_{2_{LU}}\rangle |1_{2_{RU}}\rangle \pm e^{i\eta _{U}}\epsilon
_{LU}|1_{2_{LU}}\rangle |0_{2_{RU}}\rangle \Big)  \notag \\
&&\otimes \Big(\epsilon _{RD}|0_{2_{LD}}\rangle |1_{2_{RD}}\rangle \pm
e^{i\eta _{D}}\epsilon _{LD}|1_{2_{LD}}\rangle |0_{2_{RD}}\rangle \Big)
\notag \\
&=&\epsilon _{RU}\epsilon _{RD}|0_{2_{LU}}\rangle |0_{2_{LD}}\rangle
|1_{2_{RU}}\rangle |1_{2_{RD}}\rangle \pm e^{i\eta _{U}}e^{i\eta
_{D}}\epsilon _{LU}\epsilon _{LD}|1_{2_{LU}}\rangle |1_{2_{LD}}\rangle
|0_{2_{RU}}\rangle |0_{2_{RD}}\rangle   \notag \\
&&\pm e^{i\eta _{U}}\epsilon _{RD}\epsilon _{LU}|1_{2_{LU}}\rangle
|0_{2_{LD}}\rangle |0_{2_{RU}}\rangle |1_{2_{RD}}\rangle \pm e^{i\eta
_{D}}\epsilon _{RU}\epsilon _{LD}|0_{2_{LU}}\rangle |1_{2_{LD}}\rangle
|1_{2_{RU}}\rangle |0_{2_{RD}}\rangle \text{ ,}
\end{eqnarray}
\end{widetext}%
where $\epsilon _{X}$ is the probability amplitude that a photon is
created in field $2_{X}$. The first and second terms in the
expansion correspond to the cases that the two excitations are both
retrieved at node \textquotedblleft right\textquotedblright\ and
\textquotedblleft left\textquotedblright , respectively. Thus the
effective state that yields one detection event at node
\textquotedblleft left\textquotedblright\ and the other at node
\textquotedblleft right\textquotedblright\ consists of the last two
terms. After the fields are combined by PBS$_{L}$ and PBS$_{R}$, we
get the (unnormalized) effective state of fields $2_{L}$ and $2_{R}$
\begin{equation}
|\psi _{2_{L},2_{R}}\rangle _{eff}=\alpha |H_{2_{L}}V_{2_{R}}\rangle \pm
\beta |V_{2_{L}}H_{2_{R}}\rangle \text{ ,}
\end{equation}%
where $\alpha \propto e^{i\eta _{D}}\epsilon _{RU}\epsilon _{LD}$ and $\beta
\propto e^{i\eta _{U}}\epsilon _{RD}\epsilon _{LU}$.

From the effective state $|\psi _{2_{L},2_{R}}\rangle _{eff}$, we can derive
the various coincidence probabilities $P_{ij}$, $i,j\in \{a,b,c,d\}$, where $%
\{a,b,c,d\}$ refers to the detectors $D_{2\{a,b,c,d\}}$ for field $2$ in
Fig. 1 of the main text. When $\theta _{L}$ is fixed at $0^{\circ }$, we
find (assuming unity detection efficiency)
\begin{eqnarray}
P_{ac} &=&|\alpha |^{2}\text{sin}^{2}\theta _{R}  \notag \\
P_{bd} &=&|\beta |^{2}\text{sin}^{2}\theta _{R}  \notag \\
P_{ad} &=&|\alpha |^{2}\text{cos}^{2}\theta _{R}  \notag \\
P_{bc} &=&|\beta |^{2}\text{cos}^{2}\theta _{R}  \notag \\
E(0^{\circ },\theta _{R}) &\propto &P_{ac}+P_{bd}-P_{ad}-P_{bc}=-\text{cos}%
(2\theta _{R})\nonumber\\
\end{eqnarray}%
irrespective of the $\pm $ sign.

By contrast, when $\theta _{L}$ is fixed at $45^{\circ }$, we obtain
\begin{eqnarray}
P_{ac} &=&\frac{1}{4}[1\pm 2|\alpha ||\beta |\text{ cos}\phi \text{ cos}%
(90^{\circ }-2\theta _{R})  \notag \\
&+&(|\beta |^{2}-|\alpha |^{2})\text{ sin}(90^{\circ }-2\theta _{R})]\text{ ,%
}  \notag
\end{eqnarray}%
where $\phi =arg(\beta )-arg(\alpha )$. Let $\alpha =\text{cos}\varphi $,
and \thinspace \thinspace $\beta =\text{sin}\varphi $. Denoting $\delta
=45^{\circ }-\theta _{R}$, we have
\begin{eqnarray}
P_{ac} &=&\frac{1}{4}[1\pm |\text{sin}2\varphi |\text{ cos}\phi \text{ cos}%
2\delta -\text{cos}2\varphi \text{ sin}2\delta ]  \notag \\
P_{bd} &=&\frac{1}{4}[1\pm |\text{sin}2\varphi |\text{ cos}\phi \text{ cos}%
2\delta +\text{cos}2\varphi \text{ sin}2\delta ]  \notag \\
P_{ad} &=&\frac{1}{4}[1\mp |\text{sin}2\varphi |\text{ cos}\phi \text{ cos}%
2\delta +\text{cos}2\varphi \text{ sin}2\delta ]  \notag \\
P_{bc} &=&\frac{1}{4}[1\mp |\text{sin}2\varphi |\text{ cos}\phi \text{ cos}%
2\delta -\text{cos}2\varphi \text{ sin}2\delta ]  \notag \\
E(45^{\circ },\theta _{R}) &\propto &P_{ac}+P_{bd}-P_{ad}-P_{bc}  \notag \\
&=&\pm |\text{sin}2\varphi |\text{ cos}\phi \text{ cos}2\delta \text{.}
\end{eqnarray}%
From the expression for $E(45^{\circ },\theta _{R})$, we see that the
deviation of $|\alpha |$ and $|\beta |$ from the balanced value, $1/\sqrt{2}$%
, will lead to reduction in the visibility of $E(45^{\circ },\theta _{R})$
fringes and thus the magnitudes of the CHSH parameters $S_{(\pm )}$. We
believe that such an imbalance is responsible for the results displayed in
Fig. 3(b) for $E(45^{\circ },\theta _{R})$\ and Fig. 4 for $S_{(\pm )}$ at $%
\tau _{M}=0$ in the main text, with measurements underway to quantify
this association.

Note that another combination of $P_{ij}$'s given above results in
\begin{eqnarray}
F(45^{\circ },\theta _{R}) &\equiv &-P_{ac}+P_{bd}+P_{ad}-P_{bc}  \notag \\
&=&\text{cos}2\varphi \text{ sin}2\delta \text{.}
\end{eqnarray}%
$F(45^{\circ },\theta _{R})$ allows us to determine $\varphi $ and thus the
magnitude of the coefficients $\alpha $ and $\beta $, independent of $\phi $%
. Specifically, the visibility of the $F(45^{\circ },\theta _{R})$ fringes
normalized to that of $E(0^{\circ },\theta _{R})$ fringes yields cos$%
2\varphi $.





\end{document}